\documentclass{aastex631}

\usepackage{amsmath}
\usepackage{subfigure}
\usepackage{natbib}
\usepackage{todonotes}
\usepackage{booktabs}

\usepackage{hyperref}
\hypersetup{
    colorlinks=true,
    linkcolor=blue,
    filecolor=magenta,      
    urlcolor=blue,
    pdftitle={Overleaf Example},
    pdfpagemode=FullScreen,
    }

\usepackage{dcolumn}
\usepackage{bm}

\usepackage{orcidlink}

\usepackage{graphicx}

\begin{document}

\title{The First Upper Bound on the Nano-Hertz Gravitational Waves and Galaxy Cross-Correlation signal using 15-year NANOGrav Data and DESI Galaxy Survey}

\author{Mohit Raj Sah\,\orcidlink{0009-0005-9881-1788}}
\email{mohit.sah@tifr.res.in}
\author{Suvodip Mukherjee\,\orcidlink{0000-0002-3373-5236}}
\email{suvodip.mukherjee@tifr.res.in}
\affiliation{Department of Astronomy and Astrophysics, Tata Institute of Fundamental Research, Mumbai 400005, India}

\begin{abstract}
The recent detection of a common-spectrum stochastic signal by multiple pulsar timing array (PTA) collaborations has provided tentative evidence for a nanohertz (nHz) stochastic gravitational-wave background (SGWB). This signal can be widely interpreted as originating from a cosmic population of inspiraling supermassive black hole binaries (SMBHBs). Current PTA analyses primarily constrain the SGWB power spectrum and its auto-angular power spectrum. However, the supermassive black holes will produce an underlying correlation with the large-scale structure of the Universe, which can help in understanding the formation and evolution of the binaries. In this work, we develop a new analysis pipeline \texttt{PyGxGW-PTA} for studying the cross-correlation of nHz GW signal with galaxy surveys ($C^{\rm g\, GW}_\ell$) and obtain the first constraint on the SGWB and galaxy distribution cross-correlation using the NANOGrav 15-year dataset in combination with the DESI galaxy catalog. We find no statistically significant correlation between the SGWB and the large-scale distribution of DESI galaxies and using an optimal estimator we put an upper bound on $C^{\rm g\, GW}_{\ell=8} < 0.009$ at $95\%$ C.I. This yields the first observational upper limit on the spatial correlation between the nHz SGWB and the large-scale structure of the Universe, establishing the observational groundwork for future multi-tracer analyses that will combine PTA data with next-generation galaxy surveys to unveil the SMBHB-galaxy correlation.

\end{abstract}

\section{Introduction}\label{sec:Intro}
The first evidence for a stochastic gravitational-wave background (SGWB) in the nanohertz (nHz) frequency band \citep{agazie2023nanograv,antoniadis2023second,zic2023parkes,xu2023searching} has opened a new window on the Universe.The most probable source of this signal is believed to be the inspiraling population of supermassive black hole binaries (SMBHBs). Initial PTA efforts have focused on the detection of the isotropic component of the SGWB, modeled as a power law in frequency. However, the amplitude of the SGWB reported by all PTA datasets is found to be larger than the expectations from standard theoretical and simulation-based predictions \citep{gabriela2024supermassive,agazie2023nanogravSMBHB,antoniadis2023second}. In addition, the recovered spectral slope shows deviations from the value expected for a population of circular binary evolving purely under gravitational wave (GW) radiation \citep{agazie2023nanogravSMBHB,antoniadis2023second}.

In the standard astrophysical picture, SMBHBs form following galaxy mergers and subsequently evolve under the combined action of dynamical friction, stellar scattering, torques from circumbinary gas discs, and, at the smallest separations, GW radiation reaction \citep{sampson2015constraining,kelley2017massive,chen2017efficient,taylor2017constraints,izquierdo2022massive, saeedzadeh2023shining}. The overall amplitude and spectral shape of the resulting stochastic background depend on several population level properties like the black hole–host galaxy scaling relations, the galaxy merger rate as a function of mass and redshift, the efficiency and timescale of binary hardening in different dynamical environments, and the distribution of binary orbital parameters such as total mass, mass ratio, and eccentricity \citep{feng2020supermassive,padmanabhan2023unraveling,agazie2023nanogravSMBHB,sah2024imprints, sah2025accurate, sah2025route}. However, the isotropic power spectrum measured by PTAs is, by construction, represents the integrated GW signal from the cosmic population of merging supermassive black hole binaries (SMBHBs), they carry limited information about the astrophysical origin and detailed properties of the sources contributing to the background. Therefore, the isotropic power spectrum, by construction, is degenerate with respect to several population properties. To break these degeneracies, one must exploit the spatial structure of the GW background, which encodes the clustering and anisotropies of the underlying source population \citep{sah2024imprints,sah2024discovering,sah2025route,semenzato2024cross}. The angular power spectrum of the SGWB is expected to be shot noise dominated due to the dominant contribution from a few very bright sources \citep{sah2024imprints,sato2024exploring}. In contrast, the cross-correlation between the SGWB and galaxy distributions provides a unbiased estimator of the clustering of the GW sources \citep{sah2024discovering,sah2025route}. Since SMBHBs form within galaxies, their spatial distribution should trace the large-scale matter density field. The galaxy–GW cross-correlation isolates the clustering signal between the two fields, reducing the impact of shot noise. It has been demonstrated that the amplitude and spectrum of the cross-angular power spectrum encode information about the cosmic evolution of the SMBHB population \citep{sah2024discovering,sah2025route}. Moreover, the detection (or absence) of such a correlation provides a direct diagnostic of the origin of the SGWB: a significant correlation would indicate an astrophysical origin dominated by SMBHBs, whereas a null result could point to a cosmological source such as cosmic strings, inflation, or phase transitions \citep{damour2005gravitational,siemens2007gravitational,brito2017gravitational,tsukada2019first}.

The observational prospects for measuring this cross-correlation are rapidly improving. Current galaxy surveys such as \texttt{DESI} \citep{abdul2025data} and \texttt{Euclid} \citep{aussel2025euclid}, and future wide-area programs like the \texttt{Vera C. Rubin Observatory}'s \texttt{Legacy Survey of Space and Time} (LSST) \citep{ivezic2019lsst}, will provide three-dimensional maps of the galaxy distribution extending to high redshifts. In parallel, next-generation PTAs, particularly those enabled by the \texttt{Square Kilometre Array} (SKA) \citep{ivezic2019lsst}, are expected to achieve high signal-to-noise ratio (SNR) measurements of the SGWB anisotropy. The combination of these datasets will allow the first quantitative tests of the correlation between the GW background and the large-scale structure of the Universe. Beyond the nHz band, similar methodologies can be extended to future space-based GW observatories such as \texttt{LISA} \citep{hughes2006brief,babak2011fundamental}, enabling a multi-band and multi-tracer reconstruction of the SMBHB population across cosmic time.

\begin{figure*}
    \centering    \includegraphics[width=0.8\linewidth]{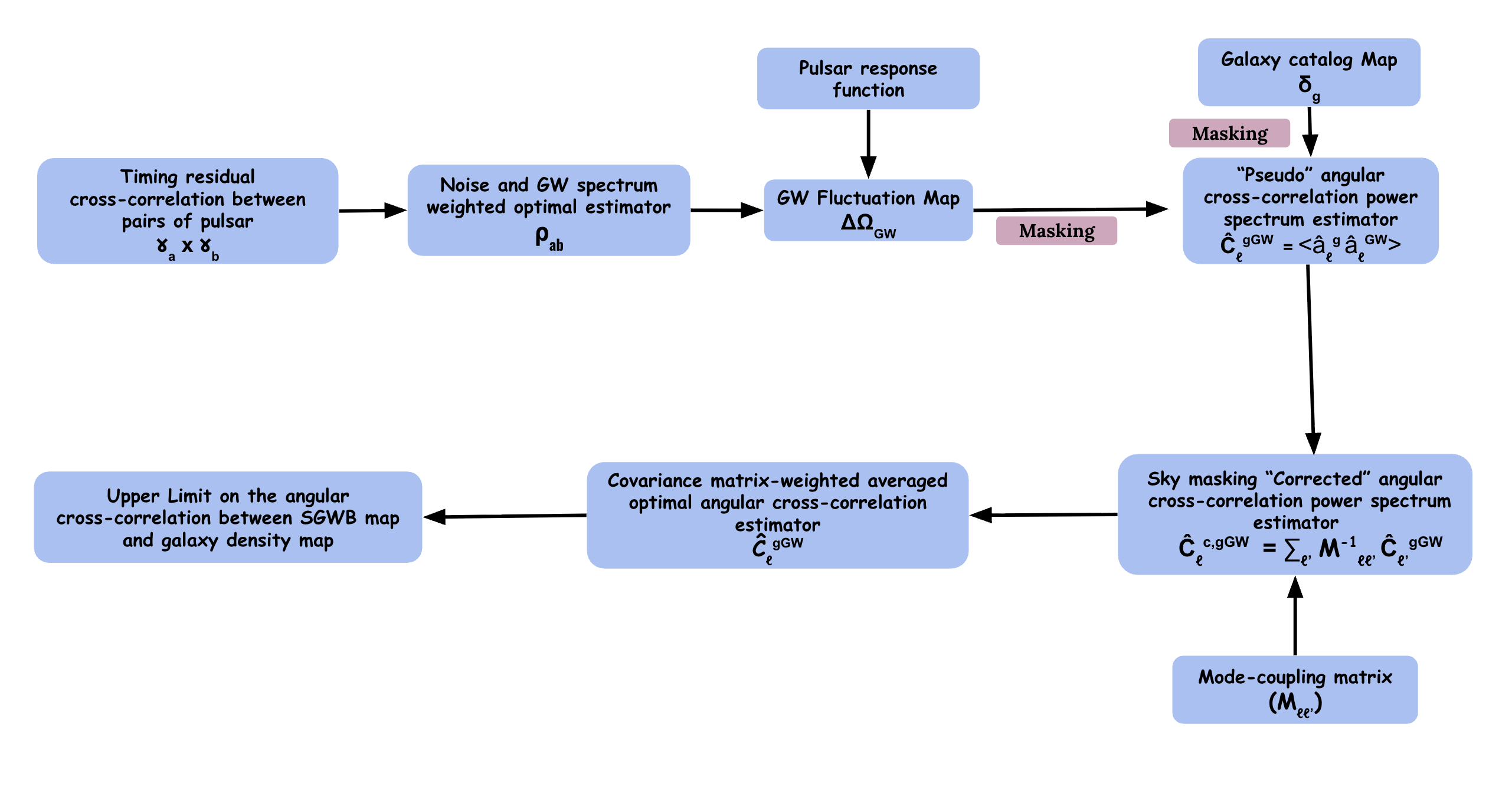}  
    \centering
     \caption{Schematic overview of the \texttt{PyGxGW-PTA} pipeline used to compute the optimal galaxy--GW cross-correlation estimator. 
    The timing-residual cross-correlations between pulsar pairs are combined to construct the SGWB map, which is then masked along with the galaxy map to compute the angular cross-correlation spectrum, $C_{\ell}^{\mathrm{gGW}}$. The "Pseudo" power spectrum is not equal to the full-sky power spectrum. The mode-coupling matrix is inverted to obtain the "mask-corrected" power spectrum. The multipole moments are optimally combined using their covariance to obtain the final estimator used for setting upper limits on the galaxy--GW correlation.}
\label{fig:flow}    
\end{figure*}

In this work, we develop the first complete analysis pipeline \texttt{PyGxGW-PTA} for measuring the galaxy–GW cross-correlation from pulsar-timing data, extending the theoretical formalism proposed in \cite{sah2024discovering} and apply its on the current PTA data.  Specifically, we apply this framework to the \texttt{NANOGrav 15-year} dataset \citep{agazie2023nanograv} in combination with the \texttt{DESI} galaxy catalog \citep{abdul2025data}. Our analysis constructs optimal estimators for the angular cross-power spectrum $C_\ell^{\mathrm{gGW}}$ to quantify the significance of the measurement. In Fig. \ref{fig:flow}, we present a schematic overview of the analysis pipeline used to compute the galaxy--GW cross-correlation estimator.  We begin by calculating the timing-residual cross-correlations for all pulsar pairs in the PTA. These pairwise measurements are then combined, using the pulsar response matrix, to reconstruct the anisotropic SGWB map. The reconstructed GW map and the galaxy density map are subsequently masked to ensure identical sky coverage before evaluating the angular cross-correlation spectrum, $C_{\ell}^{\mathrm{gGW}}$. Because masking mixes spherical-harmonic modes \citep{Hivon_2002}, the resulting pseudo--power spectrum does not directly correspond to the true full-sky power spectrum. To correct for this effect, we compute the mode-coupling matrix associated with the survey mask and invert it to obtain the mask-corrected estimate of the angular power spectrum. Finally, the individual multipoles are combined using an optimal weighting scheme based on their noise covariance, yielding the final optimal estimator of the galaxy--GW angular cross-correlation.

This analysis represents the first data-driven measurement of the galaxy–GW cross-correlation. It serves as an empirical validation of the methodology proposed by \citep{sah2024discovering} and demonstrates the practical feasibility of this approach with current PTA and galaxy survey datasets. The results presented here provide the first quantitative constraints on the spatial association between the nHz GW background and the large-scale galaxy distribution, setting the foundation for future high-precision multi-messenger analyses.  With next-generation PTAs and deeper, wider galaxy surveys, the galaxy–GW cross-correlation will provide a powerful probe of the redshift evolution of the SMBHB population \citep{sah2024discovering,sah2025route}. Because the cross-correlation amplitude and shape depends on the redshift evolution of the SMBH–host galaxy relation, the merger-rate evolution, and the mass distribution of SMBHBs, precise measurements of $C_{\ell}^{\mathrm{gGW}}(z)$ will help distinguish between different models of SMBHB evolution.

This paper is organized as follows. In Sec~\ref{sec:SGWB}, we review the SGWB and the angular galaxy–GW cross-correlation from a population of SMBHB. Sec.~\ref{sec:Residual} summarizes the construction of the timing–residual cross-correlation estimator of the PTA. In Sec.~\ref{sec:Cross}, we describe the computation of the galaxy–GW cross-correlation estimator, and \texttt{PyGxGW-PTA} reconstruction pipeline along with its validation. In Sec.~\ref{sec:Result}, we present the result of the analysis and derive upper limits on the amplitude of the galaxy–GW cross-correlation with the \texttt{NANOGrav} 15-year data combined with the \texttt{DESI} galaxy catalog.  Finally, we conclude in Sec.~\ref{sec:Conclusion} with a summary and future prospects.

\section{SGWB and the Galaxy--SGWB Cross-Correlation}\label{sec:SGWB}

The anisotropies of the SGWB encode key information about the spatial distribution of its sources and their connection to the large-scale structure of the Universe. 

This section outlines the formalism used to describe the SGWB energy density, its anisotropic fluctuations, and the resulting angular cross-correlation with galaxies. The SGWB energy density per unit logarithmic frequency and solid angle is given by \citep{phinney2001practical,christensen2018stochastic}
\begin{equation}
    \Omega_{\rm GW}(f, \hat{n}) = \frac{1}{\rho_c c^2} \int dz \frac{d^2V}{d\omega dz} \int \prod_i d\theta_i \left[ \kappa(z, \Theta_n, f_r) n_v(z, \hat{n}) \right] \left[ \frac{1}{4\pi d_L^2(z)} \frac{dE_{\rm gw}(f, \Theta_n)}{dt_r} \right],
\end{equation}
where $f_r$ is the rest-frame GW frequency, $d_L(z)$ is the luminosity distance at redshift $z$, $\omega$ is the solid angle. The integration extends over the comoving volume element $dV$, and the binary-parameter set $\Theta_{\rm n} = \{\theta_i\}_{i=1}^{n}$, which includes quantities such as component masses and orbital inclination. The term ${dE_{\rm gw}(f_r, \Theta_{\rm n})}/{dt_r}$ denotes the GW luminosity of a binary characterized by $\Theta_{\rm n}$, while $n_v(z, \hat{n})$ is the number of galaxies per unit comoving volume. The function $\kappa(\Theta_{\rm n}, f_r \mid z)$ represents the mean number of SMBHBs per galaxy at redshift $z$, with parameters $\Theta_{\rm n}$.

The anisotropic fluctuations in the SGWB can be defined as \cite{sah2024discovering,sah2025route}
\begin{equation}
    \begin{aligned}
    \delta_{\rm GW}(\hat{n}) \equiv  & {\frac{ \int \frac{df}{f} ~  \Omega_{\rm GW}(f,\hat{n}) - \frac{df}{f}  ~ \overline{\Omega}_{\rm GW}(f)}{\int \frac{df}{f} ~\overline{\Omega}_{\rm GW}(f)}},\\
    = &\int\limits_{{\rm z}_{\rm min}}^{\infty} ~d\rm{ z} ~  \phi_{\rm GW}(\rm{z}) ~ \delta_{\rm m}({\rm z},\hat{n}) ~ b_{\rm GW}(z), 
    \end{aligned}
    \label{SGWB_Fluc}
\end{equation}
where $\overline{\Omega}_{\rm GW}(f)$ is the mean (sky-averaged) SGWB energy density. The quantity $\delta_{\rm m}({\rm z}, \hat{n})$ represents the matter overdensity field, and $b_{\rm GW}({\rm z})$ is the effective GW bias that quantifies how GW sources trace the underlying matter distribution. The function $\phi_{\rm GW}({\rm z})$  acts as a redshift-dependent weighting function determining how different redshifts contribute to the SGWB. It is defined as

\begin{equation}
    \phi_{\rm GW}(z)  = \frac{\frac{dn}{d{\rm z}}  \int  \frac{df}{f} ~  \int \prod\limits_{i}^{n} \frac{d\theta_i}{ d_{\ell}^{2} } \Big[\overline{\kappa}({\rm z},\Theta_{\rm n},f_{\rm r})\Big] \Big[\frac{ dE_{\rm gw}(f_{{\rm r}},\Theta_{\rm n})}{dt_{\rm r}}\Big]}{\int\limits_{{\rm z}_{\rm min}}^{\infty} \frac{d\rm{z}}{d_{\ell}^{2}} \frac{dn}{d\rm{z}} \int \frac{df}{f} \int  \prod\limits_{i}^{n} d\theta_i ~    \Big[\overline{\kappa}({\rm z},\Theta_{\rm n},f_{\rm r})\Big] \Big[\frac{ dE_{\rm gw}(f_{{\rm r}},\Theta_{\rm n})}{{ dt_{\rm r}}}\Big]},
    \label{phiGW}
\end{equation}
where ${dn}/{d{\rm z}}$ denotes the number of galaxies per unit redshift bin.

The fluctuations in the SGWB ($\delta_{\rm GW}(\hat{n})$) and in the galaxy density field ($\delta_{\rm g}(\hat{n})$) can be expanded in spherical harmonics as
\begin{align}
    \delta_{\rm g,GW}(\hat{n}) =& \sum\limits_{\ell} \sum\limits_{m = -\ell}^{\ell}  a_{\ell m}^{\rm g,GW} Y_{\ell m}(\hat{n}),
\end{align}
and their angular cross-correlation power spectrum is given by \citep{sah2024discovering,sah2025route}
\begin{equation}
    C_\ell^{g\,{\rm GW}} = \left\langle a_{\ell m}^{g} a_{\ell m}^{\rm GW\,*} \right\rangle = \frac{2}{\pi} \int dz_1\, W_g(z_1) D(z_1) \int dz_2\, W_{\rm GW}(z_2) D(z_2) \int k^2 dk\, j_\ell(k r_1) j_\ell(k r_2) P(k),
\end{equation}
where $W_g(z) = b_g(z)\,\phi_g(z)$ and $W_{\rm GW}(z) = b_{\rm GW}(z)\,\phi_{\rm GW}(z)$ are the window functions for galaxies and GWs, respectively, $D(z)$ is the linear growth factor, and $P(k)$ is the matter power spectrum. The spherical Bessel functions $j_\ell(kr)$ project the three-dimensional correlations onto the celestial sphere.

\section{Timing Residual Cross-Correlation Estimator}\label{sec:Residual}

A GW passing between the Earth and a pulsar perturbs the pulse arrival times, producing a deviation known as a timing residual, $r(t)$. The cross-correlation of the timing residuals from two pulsars provides a direct measure of the SGWB-induced correlations. 

%A detailed derivation of the timing residual response is given in Appendix~\ref{sec:timing_crosscorr}.

For each pulsar pair, we construct an optimal cross-correlation estimator by combining all timing-residual measurements in a way that maximizes the SNR. We denote the optimal estimator for pulsars $a$ and $b$ by $\rho_{ab}$, with an associated uncertainty $\sigma_{ab}$ \citep{anholm2009optimal,siemens2013stochastic,pol2022forecasting,agazie2023nanograv}. These quantities are defined as
\begin{align}
    \rho_{ab} &=
    \frac{r_a^{T} P_a^{-1} \hat{S}_{ab} P_b^{-1} r_b}         {\mathrm{tr}\!\left[P_a^{-1} \hat{S}_{ab} P_b^{-1} \hat{S}_{ba}\right]},
    \label{eq:rho} \\
    \sigma_{ab} &=
    \left[\,\mathrm{tr}\!\left(P_a^{-1} \hat{S}_{ab} P_b^{-1} \hat{S}_{ba}\right)\right]^{-1/2},
    \label{eq:sigma}
\end{align}
where $r_a$ and $r_b$ are the timing-residual vectors, and $P_a$ represents the auto-covariance matrix
of the timing residual of pulsar a. The matrix $\hat{S}_{ab}$ represents the GW-induced covariance between the two pulsars. It encodes the spectral dependence of the expected SGWB signal (e.g., $\propto f^{-13/3}$ for circular SMBHBs) and the angular correlation pattern across the sky. In the present analysis, we adopt a GW-induced cross-correlation spectrum with spectral index $-7/2$, corresponding to the median value reported by the NANOGrav 15-year dataset.

Once the optimal timing residual cross-correlations $\rho_{ab}$ have been measured, the next step is to infer the SGWB map. This involves estimating the GW power map $P(\hat{\Omega})$ that best explains the observed cross-correlations. The likelihood function for the cross-correlation data $\boldsymbol{\rho}$ can be written as

\begin{equation}
    \mathcal{L}(\boldsymbol{\rho} | \mathbf{P}) \propto \exp\left[ -\frac{1}{2} (\boldsymbol{\rho} - \mathbf{R} \mathbf{P})^T \boldsymbol{\Sigma}^{-1} (\boldsymbol{\rho} - \mathbf{R} \mathbf{P}) \right],
    \label{eq:Like}
\end{equation}
where $\boldsymbol{\Gamma} = \mathbf{R}\,\mathbf{P}$ is the overlap reduction function (ORF), with $\mathbf{P}$ denoting the vector of GW power either in pixel space or in spherical-harmonic coefficients, depending on the chosen basis. The matrix $\mathbf{R}$ is the overlap response matrix constructed from the pulsar antenna patterns, and $\boldsymbol{\Sigma}$ is the covariance matrix of the timing-residual cross-correlation measurements, which is diagonal with elements $\sigma_{ab}^2$.

We adopt the spherical harmonic basis, in which the GW power is decomposed as
\begin{equation}
    P(\hat{n}) = \sum_{\ell m} a_{\ell m} Y_{\ell m}(\hat{n}),
\end{equation}
where $a_{\ell m}$ is the coefficients of the spherical harmonic basis vector $Y_{\ell m}(\hat{n})$. The ORF can now be written as 
\begin{equation}
    \begin{aligned}
    \boldsymbol{\Gamma}_{ab} \propto &
     \sum_{\ell, m} c_{lm} \sum_{k} Y_{lm}(\hat{n}_k)
    \left[
    \mathcal{F}^{+}_{a}(\hat{n}_k) \mathcal{F}^{+}_{b}(\hat{n}_k)
    + 
    \mathcal{F}^{\times}_{a}(\hat{n}_k) \mathcal{F}^{\times}_{b}(\hat{n}_k)
    \right],\\
    & = \mathbf{P}_{\rm S} \mathbf{R}_{\rm S},
    \end{aligned}
    \label{eq:Gamma_lm}
\end{equation}
where $k$ denotes the pixel index on the sky, $\mathbf{P}_{\rm S}$ is the vector containing the spherical harmonic coefficients of the SGWB density field, and $\mathbf{R}_{\rm S}$ represents the corresponding PTA response matrix for the spherical harmonic basis. For brevity, we omit the subscript ``S'' in the following discussion.

For basis choices where $\mathbf{P}$ enters linearly in the model (e.g., the pixel basis or the spherical harmonic basis), the likelihood can be analytically maximized to yield the following maximum-likelihood estimator

\begin{equation}
    \mathbf{P}_{\rm ML} = (\mathbf{R}^T \boldsymbol{\Sigma}^{-1} \mathbf{R})^{-1} \mathbf{R}^T \boldsymbol{\Sigma}^{-1} \boldsymbol{\rho}. 
    \label{eq:linear_ml}
\end{equation}

This solution is linear in $\boldsymbol{\rho}$ and the covariance of the estimator is given by the inverse Fisher matrix
\begin{equation}
    \mathrm{Cov}(\mathbf{P}) = (\mathbf{R}^T \boldsymbol{\Sigma}^{-1} \mathbf{R})^{-1}.
\end{equation}

\section{Galaxy--GW Cross-Power Spectrum Estimator}\label{sec:Cross}

In this section, we describe our pipeline \texttt{PyGxGW-PTA} in detail explaining the construction of the estimator, the treatment of partial sky coverage, and the noise properties of the reconstructed GW map. We also validate our pipeline using mock realizations before applying it to real data.

\subsection{Reconstructed GW Map and Harmonic Coefficients}
The cross-correlation analysis requires both the galaxy density map and the reconstructed GW anisotropy map to be represented in a common spherical-harmonic basis. The galaxy field ($\delta_{\rm g}(\hat{n})$) is decomposed as $\hat{a}_{\ell m}^{g}$, and the SGWB map ($\delta_{\rm GW}(\hat{n})$) yields harmonic coefficients $\hat{a}_{\ell m}^{\mathrm{GW}}$ obtained from the maximum-likelihood reconstruction from Eq. \eqref{eq:linear_ml}. The cross-power spectrum is then computed as
\begin{equation}
    \hat{C}_{\ell}^{g\,\mathrm{GW}} =
    \frac{1}{2\ell + 1}
    \sum_{m=-\ell}^{\ell}
    \hat{a}_{\ell m}^{g}
    \hat{a}_{\ell m}^{\mathrm{GW} *}.
    \label{eq:cross_spectrum_estimator}
\end{equation}

\subsection{Partial Sky Coverage and Mode Coupling}

The spherical-harmonic transform of a field on a sky map with incomplete sky coverage, $\hat{C}^{g\,\mathrm{GW}}_\ell$ (pseudo–power spectrum), is not equal to the true full-sky power spectrum $C^{g\,\mathrm{GW}}_\ell$. However, their ensemble averages are connected through a linear relation \citep{Hivon_2002}
\begin{equation}
    \langle \hat{C}^{g\,\mathrm{GW}}_\ell \rangle = \sum_{\ell'} M_{\ell\ell'} \, \langle C^{g\,\mathrm{GW}}_{\ell'} \rangle,
\end{equation}
where the matrix $M_{\ell\ell'}$ is the mode-coupling matrix, which quantifies the coupling between multipoles that arises due to the sky mask. The coupling matrix is given by \citep{Hivon_2002}
\begin{equation}
    M_{\ell \ell'} =
    \frac{2\ell' + 1}{4\pi}
    \sum_{\ell''} (2\ell'' + 1)
    W_{\ell''}
    \begin{pmatrix}
        \ell & \ell' & \ell'' \\
        0 & 0 & 0
    \end{pmatrix}^{2},
\label{eq:Mll}
\end{equation}
where $W_\ell$ is the power spectrum of the mask. For a given mask $W(\hat{n})$, its power spectrum is given by
\begin{equation}
    W_\ell = \frac{1}{2\ell + 1}
    \sum_{m=-\ell}^{\ell} 
    \left| w_{\ell m} \right|^{2},
\label{eq:Wl}
\end{equation}
with $w_{\ell m}$ being the spherical-harmonic coefficients of the mask. We invert the mode-coupling matrix to obtain the mask-corrected spectrum, $\hat{C}_{\ell}^{c, g\,\mathrm{GW}}$.
\begin{equation}
    \hat{C}_{\ell}^{c,g\,\mathrm{GW}}
    = \sum_{\ell'} M^{-1}_{\ell\ell'} \,
    \hat{C}_{\ell'}^{g\,\mathrm{GW}},
\end{equation}

The matrix $\hat{C}_{\ell}^{c,g\,\mathrm{GW}}$ gives the corrected power spectrum for the masked sky.

\subsection{Noise Properties and Anisotropic Covariance}

The reconstructed GW map inherits anisotropic noise arising from both the non-uniform angular distribution of PTA pulsars on the sky and the pulsar-dependent noise properties encoded in their individual timing-residual covariance matrices \citep{agazie2023nanograv_noise}. In addition, the incomplete and highly irregular sky coverage of the galaxy catalog  leads to a non-trivial mode coupling between spherical-harmonic coefficients of the reconstructed GW field. As a result, different angular modes of the GW map are no longer statistically independent, and the noise covariance becomes intrinsically anisotropic and non-diagonal in harmonic space. Accurately accounting for this covariance is therefore essential for unbiased estimation of the galaxy--GW cross-correlation signal and its statistical significance.

\begin{figure*}
    \centering
    \includegraphics[width=0.7\linewidth]{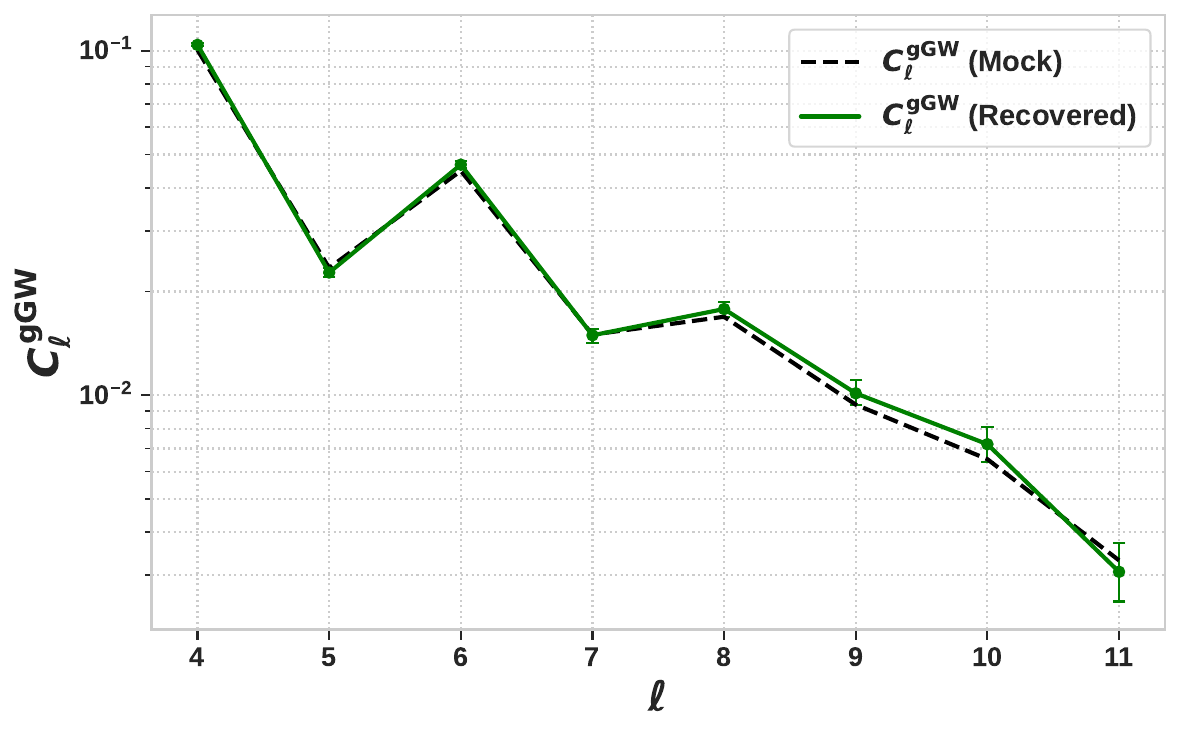}
    \caption{Validation of the cross-correlation pipeline using mock data.  
    The dashed black curve shows the injected angular cross-power spectrum, $C_{\ell}^{\rm gGW}\,(\mathrm{Mock})$, constructed by forcing the SGWB anisotropy to follow the DESI galaxy density fluctuations.  
    The solid green curve shows the recovered spectrum obtained after passing the noisy mock timing-residual cross-correlations through the full reconstruction pipeline, with error bars estimated from $\sim 1000$ independent realizations. We simulate 200 isotropically distributed pulsars with the timing-residual noise level set to be 100 times smaller than the median uncertainty of the NANOGrav pulsars to illustrate the recovery of the method. The close agreement between the injected and recovered spectra demonstrates that the estimator reliably reconstructs the underlying galaxy--GW correlation.}
    \label{fig:CgGW_valid}
\end{figure*}

\subsection{Validation}
To test the performance of our cross-correlation pipeline in a controlled setting, we construct mock datasets through the following steps

%Using this mock GW map and the pulsar response functions, we compute the cross-correlation signal $\rho^{\rm mock}_{ab}$ for each pulsar pair.

\begin{itemize}
    \item  We start with the observed \texttt{DESI} galaxy map and compute the galaxy overdensity field, $\delta_g(\hat{n})$.

    \item The mock SGWB fluctuation field ($\delta_{\rm GW}^{\rm mock}(\hat{n})$) is constructed to follow the galaxy density fluctuation map, $\delta_{\rm GW}^{\rm mock}(\hat{n}) \propto \delta_g(\hat{n})$ ensuring a non-zero and controlled galaxy--GW cross-correlation. The overall amplitude is normalized to match the median value of the power-law spectrum inferred from the \texttt{NANOGrav} 15-year dataset \citep{agazie2023nanograv}.

    \item  Using this mock SGWB map and the pulsar antenna response functions, we compute the cross-correlation signal $\rho^{\rm mock}_{ab}$ for each pulsar pair.

    \item Each pulsar pair is assigned a mock timing-residual cross-correlation uncertainty, $\sigma^{\rm mock}_{ab} = \alpha\, \sigma^{\rm \rm median}_{ab}$, where $\sigma^{\rm \rm median}_{ab}$ is the median timing residual cross-correlation measurement uncertainity of the NANOGrav pulsars, and  $\alpha$ is a constant. 
 
    \item The measured mock correlations are generated by sampling from a Gaussian distribution, $\rho^{\rm obs}_{ab} \sim 
        \mathcal{N}\!\left(\rho^{\rm mock}_{ab},\, \sigma^{\rm mock}_{ab}\right)$, where $\mathcal{N}$ represents the normal distribution. 
\end{itemize}

\begin{itemize}
    \item This ensemble of noisy $\rho^{\rm obs}_{ab}$ values is then processed through the reconstruction pipeline to obtain the maximum likelihood estimate of the SGWB map, using \textsc{Healpy} \citep{2005ApJ...622..759G, Zonca2019} with a pixel resolution parameter of \texttt{nside} = 4 (corresponding to a maximum multipole moment of $\ell_{\mathrm{max}} = 11$), as described in the previous subsections (Eq. \eqref{eq:linear_ml}).
    
    \item We mask both the SGWB map and the galaxy density fluctuation map for the sky coverage of the DESI catalog.
    
    \item  The masked map is then cross-correlated to obtain the power spectrum ($C_{\ell}^{\rm gGW}$).
\end{itemize}

In Fig. \ref{fig:CgGW_valid}, we show an example of recovering the $C_{\ell}^{\rm gGW}$ under ideal conditions described above. We assume 200 isotropically distributed pulsars, and the timing residual cross-correlation uncertainty for all pulsar pairs is 100 times smaller than that of the median values of the \texttt{NANOGrav} pulsars (i.e, $\alpha = 10^{-2}$). We see that the algorithm is able to reproduce the injected $C_{\ell}^{\rm gGW}$ spectrum fairly well, demonstrating that the method can accurately recover the underlying SGWB anisotropy.

To further test whether the reconstruction introduces any systematic bias, we perform an additional validation in which both the galaxy map and the SGWB map are replaced by independent random Gaussian fields, masked identically to the \texttt{DESI} sky coverage. All other analysis settings are kept unchanged. For each realization, we run the full reconstruction and cross-correlation pipeline and compute the resulting $C_{\ell}^{\rm gGW}$.

\begin{figure*}
    \centering
    \includegraphics[width=0.7\linewidth]{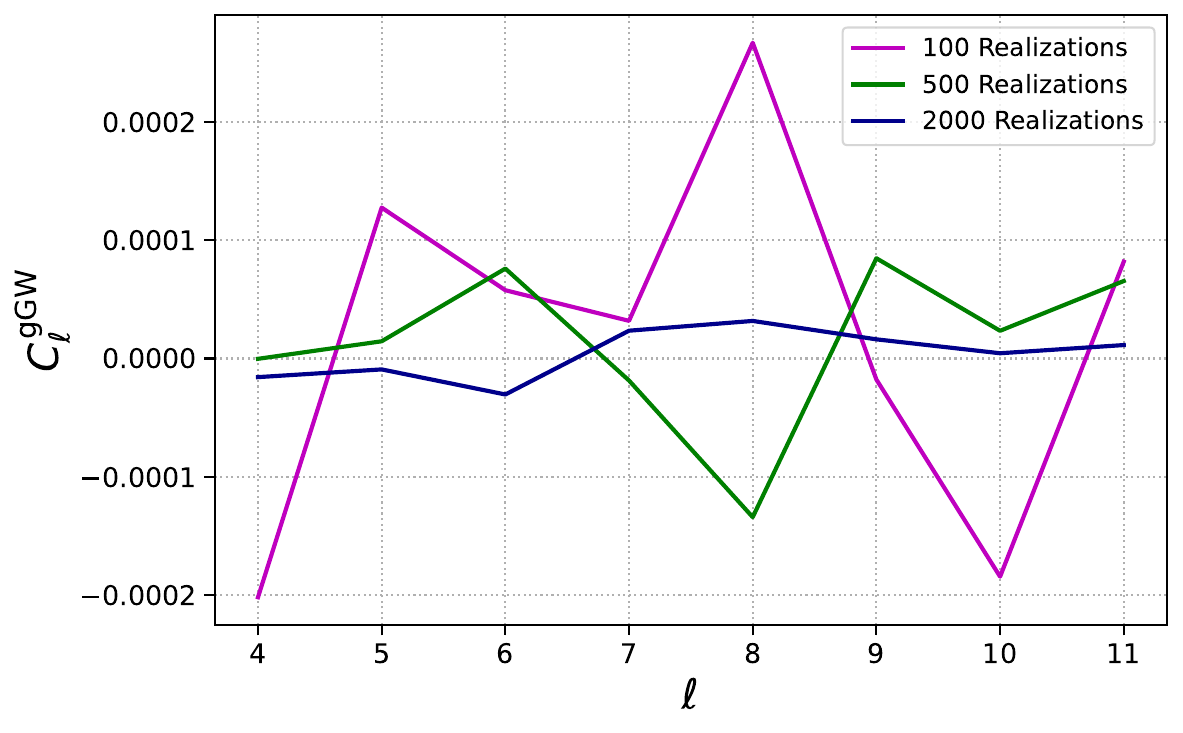}
    \caption{Ensemble-averaged galaxy–GW cross-power spectra obtained from random, statistically independent Gaussian realizations of both the SGWB anisotropy map and the galaxy density map. All maps are masked with the DESI sky coverage, and the reconstruction settings match those used in Fig.~\ref{fig:CgGW_valid}. The curves show the mean recovered spectrum averaged over 100, 500, and 2000 realizations. As the number of realizations increases, the mean $C_{\ell}^{\rm gGW}$ approaches zero across all multipoles, indicating that the pipeline does not introduce spurious correlations from masking, noise in the map reconstruction, or pulsar geometry. This confirms that the estimator is statistically unbiased in the absence of a true galaxy–GW correlation signal.}
    \label{fig:CgGW_random}
\end{figure*}

In Fig.~\ref{fig:CgGW_random}, we show the mean recovered cross-correlation spectrum obtained by averaging over 100, 500, and 2000 realizations. As the number of realizations increases, the ensemble-averaged spectrum converges rapidly toward zero across all multipoles, confirming that the estimator is unbiased in the absence of a true correlation, and the masking and map-making procedure do not introduce artificial cross-power.

\begin{figure*}
    \centering    \includegraphics[width=0.8\linewidth]{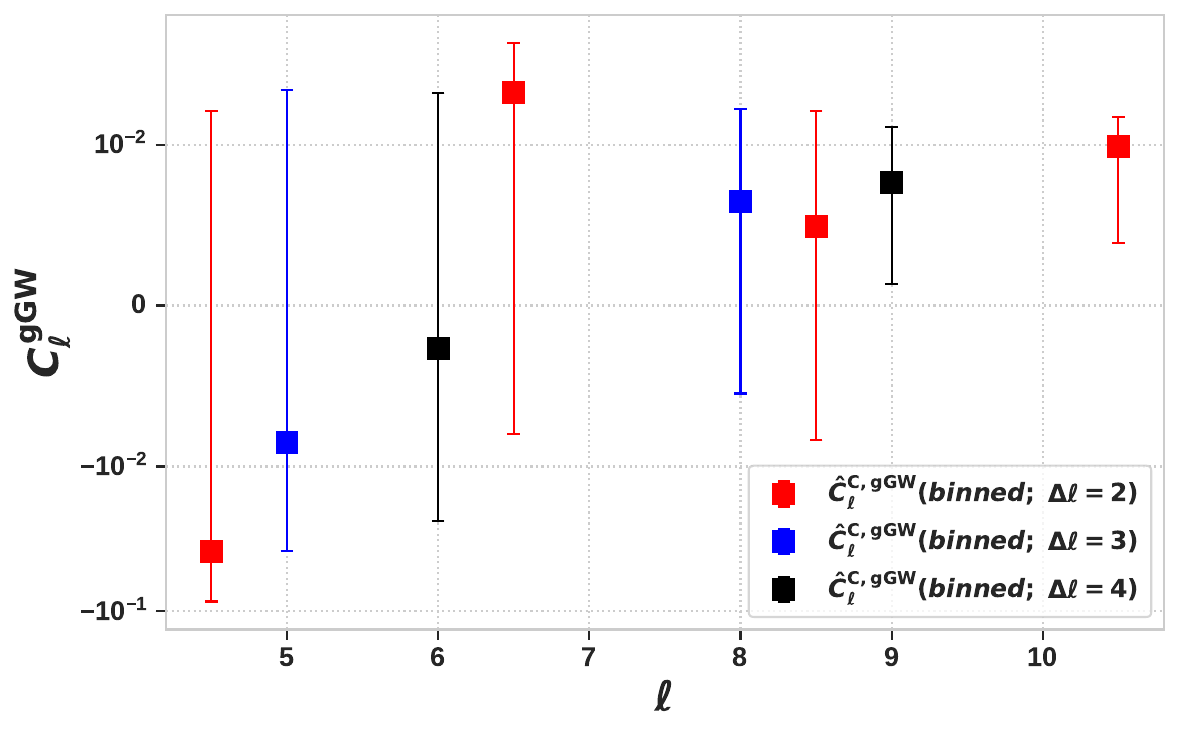}  
    \centering
    \caption{Angular power spectra of the galaxy–GW cross-correlation obtained using the \texttt{DESI} galaxy catalog and the \texttt{NANOGrav} 15-year dataset. The square markers with error bars show the masking-corrected estimates of $\hat{C}_\ell^{\mathrm{C,gGW}}$, after averaging over multipole bins of width $\Delta\ell = 2$, $3$, and $4$. The error bars represent the uncertainties in the measured cross-correlation, derived from Monte Carlo realizations of $C_\ell^{\mathrm{C,gGW}}$ under the null hypothesis of no correlation.}
\label{fig:CgGW}    
\end{figure*}

\section{Results and discussion}\label{sec:Result}

\begin{figure*}
    \centering    \includegraphics[width=0.65\linewidth]{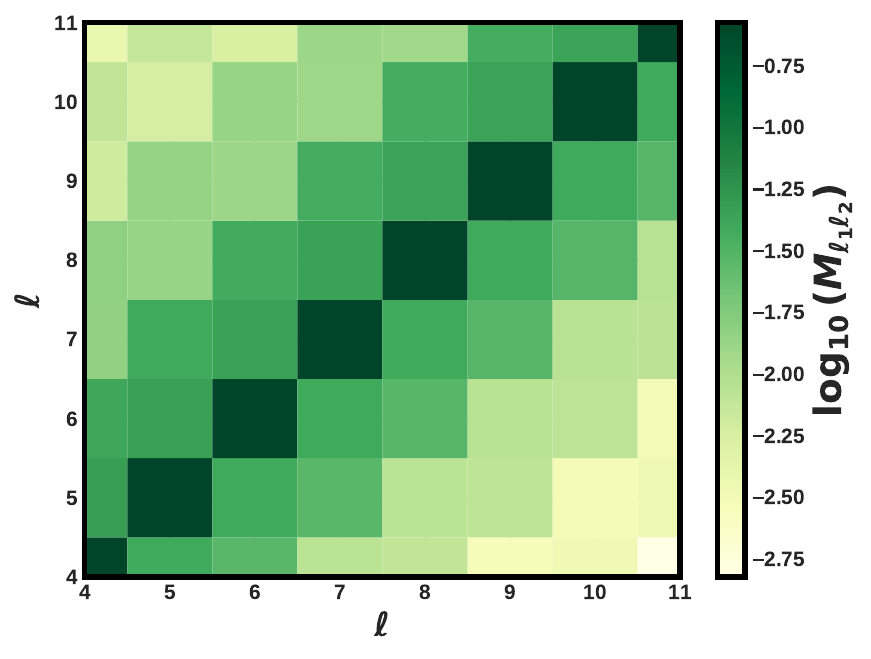}  
    \centering
    \caption{Logarithm of the mode–coupling matrix, $\log_{10}(M_{\ell_1\ell_2})$, induced by the sky mask. The matrix is computed from the window function of the DESI galaxy catalog, and the elements quantify how the true sky power at multipole $\ell_1$ leaks into the multipole $\ell_2$. The presence of off–diagonal structure encodes leakage introduced by incomplete sky coverage. The color scale represents the logarithmic amplitude of the coupling coefficients.}
\label{fig:Mll}    
\end{figure*}

\begin{figure*}
    \centering    \includegraphics[width=0.65\linewidth]{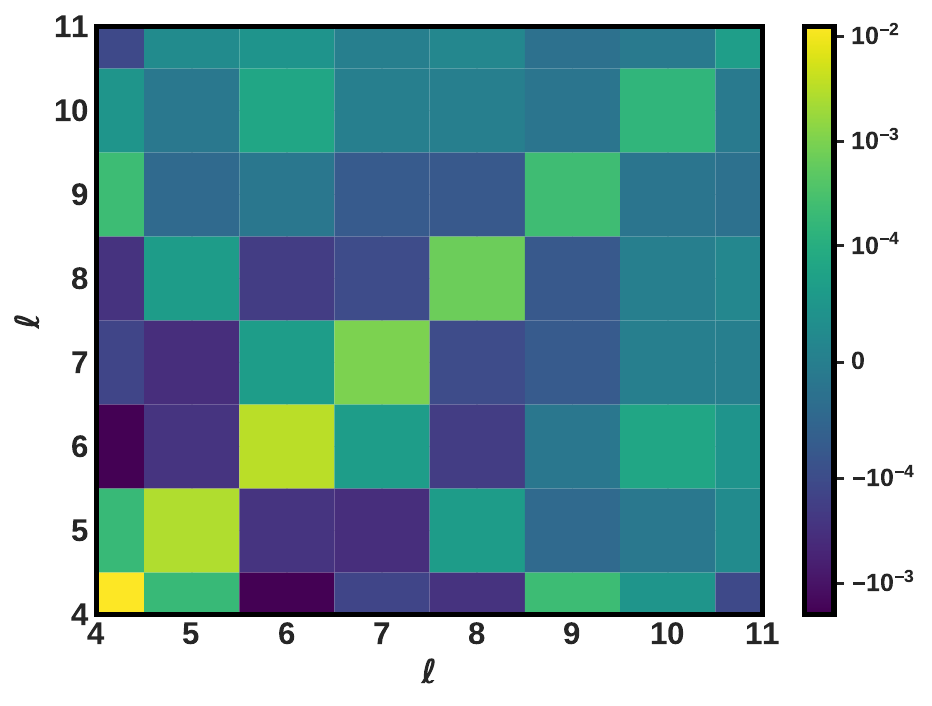}  
    \centering
    \caption{Covariance matrix of the cross-correlation estimator $\hat{C}_\ell^{g\,\mathrm{GW}}$ between different multipole moments, obtained from multiple noise realizations. The color scale represents the degree of statistical correlation between multipoles. The visible off-diagonal structure arises from incomplete sky coverage, anisotropic pulsar distribution with non-uniform noise properties.}
\label{fig:Cov}    
\end{figure*}

\subsection{First Cross-Correlation Inference Using \texttt{NANOGrav} Observations and \texttt{DESI} Galaxies}

We apply our formalism to the \texttt{NANOGrav 15-year} dataset \citep{agazie2023nanograv} in combination with the \texttt{DESI} galaxy catalog \citep{abdul2025data}. In Fig. \ref{fig:CgGW}, we show the measured galaxy–GW cross-angular power spectrum, $\hat{C}_{\ell}^{c,\mathrm{gGW}}$ obtained with the \textsc{Healpy} resolution of \texttt{nside} = 4. The $\hat{C}_{\ell}^{c,\mathrm{gGW}}$ quantifies the degree of spatial correlation between the SGWB anisotropy reconstructed from the \texttt{NANOGrav} 15-year dataset and DESI galaxies. The error bars correspond to the standard deviation obtained from 1000 Monte Carlo realizations generated under the null hypothesis of no galaxy–GW correlation. The square markers show the masking-corrected estimates of cross-correlation, averaged over multipole bins of width 
$\Delta\ell = 2$, $3$, and $4$. 

In Fig. \ref{fig:Mll} we show the logarithm of the coupling matrix (defined in Eq. \eqref{eq:Mll}, $\log_{10}(M_{\ell_1\ell_2})$), restricted to the range of multipoles used in our cross-correlation analysis. This matrix characterizes how the true underlying sky power at multipole $\ell_1$ is redistributed into the multipole $\ell_2$ due to the incomplete and non-uniform sky coverage of the galaxy sample. The matrix has notable off-diagonal features, reflecting leakage between different multipoles induced by the geometry of the mask. These off-diagonal couplings are accounted for by inverting $M_{\ell_1\ell_2}$ when reconstructing the unbiased sky angular power spectra. After taking into account the mode-coupling due to masking, we obtain the cross-correlation covariance matrix shown in Fig. \ref{fig:Cov}, evaluated across multiple realizations of timing residual noise and galaxy catalog.

\subsection{Optimal detection statistic for the angular cross-correlation}
We define an optimal detection statistic for the angular cross-correlation between the galaxy overdensity field and the SGWB anisotropy by constructing a weighted combination of the cross-power spectrum multipoles, $\hat{C}_\ell^{c,g\,\mathrm{GW}}$. The purpose of this weighting is to maximize the SNR ratio by accounting for both the noise covariance and the intrinsic clustering of the galaxy field. This leads to the generalized optimal estimator \footnote{Derivation in Appendix~\ref{sec:Der_opt}.}

\begin{equation}
    \mathcal{\hat{C}}^{g\,\mathrm{GW}} \equiv (\hat{\mathbf{C}}^{g\,\mathrm{GW}})^{\mathrm{T}}\,
    \Sigma_{\mathrm{gGW}}^{-1}\, \mathbf{C}_t^{\mathrm{gGW}}~\Bigg[\frac{\mathbf{C}_t^{gGW}(\ell_{\rm ref})}
    {(\mathbf{C}_t^{gGW})^{\mathrm{T}}\,
    \Sigma_{\mathrm{gGW}}^{-1}\,
    \mathbf{C}_t^{\mathrm{gGW}}}\Bigg],
    \label{eq:opt_corr_estimator}
\end{equation}

where $\hat{\mathbf{C}}^{g\,\mathrm{GW}}$ is the vector of cross-power spectrum multipoles $\hat{C}_{\ell}^{c,g\,\mathrm{GW}}$, $\Sigma_{\mathrm{gGW}}$ is the covariance matrix of the estimator computed from Monte Carlo realizations under the null hypothesis, and $\mathbf{C}_t^{\mathrm{gGW}}$ denotes the theoretical template of cross-correlation power spectrum. The quantity in the square bracket is a normalization that is choosen such that the estimator represents the angular cross-correlation at a reference $\ell$ ($\ell_{\rm ref}$). The amplitude and the shape of the cross-correlation spectrum, $C_{\ell}^{\rm gGW}$,  depend primarily on the redshift evolution of the population \citep{sah2024discovering,sah2025route}. For the theoretical $C_{\ell}^{\rm gGW}$, we assume that the redshift evolution of the number density of the SMBHB follows the galaxy number density of the galaxy catalog.

In addition, we assume that all binaries are on circular orbits, with their orbital-frequency distribution following the standard power-law expected for a population of inspiraling circular binary (i.e; $P(f) \propto f^{-11/3}$). This simplification ignores the fact that, in reality, SMBH binaries are expected to retain significant eccentricity in the GW-emission regime \citep{chen2017efficient, gualandris2022eccentricity}. Such residual eccentricity would modify the spectral shape of $\Omega_{\rm GW}(f)$ and, in turn, affect the galaxy–GW cross-correlation power spectrum \citep{sah2024discovering, sah2025route}. However, this simplification is adequate for our purposes, as the current data are not yet sensitive to the subtle changes induced by eccentricity, making the circular-binary model a sufficiently robust choice.

We construct the template, $\mathbf{C}_{t}^{\rm gGW}(\ell_{\rm ref})$, from a physically motivated population of SMBHB. We assume a linear scaling relation between the SMBH mass and the stellar mass of the host galaxy, given by \citep{reines2015relations,habouzit2021supermassive,kozhikkal2024mass} 

\begin{equation}
    \mathrm{Log}_{10}(M_{\rm BH}/M_{\odot}) = \eta + \rho ~ \mathrm{Log}_{10}( M_{*}/10^{11} M_{\odot}) + \nu  ~z, 
    \label{MBH1}
\end{equation}
where $ \mathrm{Log}_{10}(M_{\rm BH})$ is the SMBH mass and $ M_{*}$ is the stellar mass of the host galaxy. $\eta$, $\rho$, and $\nu$ are free parameters governing the scaling relation and its redshift evolution. The detailed model of the SMBHB population is presented in the Appendix \ref{sec:pop}.
For this analysis, we assume $\eta = 8$, $\rho = 1$. These values are motivated by the various simulations and observational studies of the $M_{*}-M_{\rm BH}$ relation for both isolated and binary SMBHs \citep{reines2015relations,habouzit2021supermassive,saeedzadeh2023shining,kozhikkal2024mass}. The redshift evolution of this relation is not well constrained, so we consider results for four representative choices of the evolutionary parameter, $\nu$ = -2, -0.5, 0, and 0.5. A positive value of $\nu$ corresponds to a scenario in which SMBHs are more massive at fixed host-galaxy stellar mass at higher redshifts, while a negative value of $\nu$ implies the opposite trend—namely, that SMBHs are comparatively lighter in galaxies of the same stellar mass at higher redshifts.

\subsection{Upper limit on the galaxy-GW cross-correlation}

\begin{figure*}
    \centering    \includegraphics[width=0.8\linewidth]{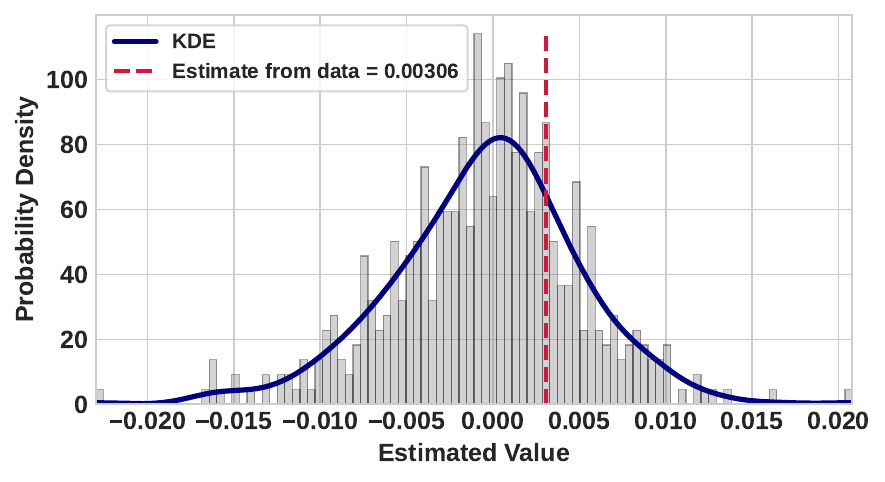}  
    \centering
    \caption{
    Kernel density estimate (KDE) of the cross-correlation estimator distribution obtained from Monte Carlo realizations (for $\nu =0$ template model). The red vertical line indicates the optimal estimate of the statistic (= 0.00306), while the gray histogram shows the underlying sample distribution. The $p$-value ($\approx 0.23$) is computed as the fraction of realizations exceeding the observed statistic, quantifying the statistical significance of the measured galaxy–GW cross-correlation.}
\label{fig:p-value}    
\end{figure*}

In the absence of a signal, random noise fluctuations do not align with the galaxy clustering pattern, and the statistic remains consistent with zero. In this way, the estimator isolates the component of the cross-correlation that is physically linked to the large-scale structure traced by galaxies. The estimator defined in Eq.~\eqref{eq:opt_corr_estimator} optimally combines the modes by down-weighting modes dominated by noise, partial-sky coupling, or sample variance. 

In Fig.~\ref{fig:p-value}, we present the kernel density estimate (KDE) of the cross-correlation statistic obtained from Monte Carlo realizations for $\nu = 0$. The red vertical line marks the measured value from the data, while the gray histogram shows the distribution of the estimator under the null hypothesis. In the absence of a statistically significant detection of the galaxy--GW correlation, we derive an upper limit on its amplitude. To do so, we generate a large ensemble of simulated realizations of $\hat{\mathcal{C}}^{g\,\mathrm{GW}}$ under the null hypothesis (no GW signal) using the noise properties of the \texttt{NANOGrav} pulsar array. The empirical distribution of $\hat{\mathcal{C}}^{g\,\mathrm{GW}}$ is then used to compute the $p$-value of the value measured from the data. We obtain a $p$-value of approximately $ 0.23$, indicating that the observed cross-correlation is fully consistent with statistical fluctuations expected in the absence of any correlated SGWB signal. Consequently, we place an upper limit on the amplitude of the galaxy--GW cross-correlation (  $\ell_{\rm ref} = 8$) at the $95\%$ confidence level, which we denote as $\mathcal{C}^{g\,\mathrm{GW}}_{\mathrm{UL}}  \leq 0.009$. 

The $p$-values obtained for each of the theoretical galaxy–GW correlation templates are summarized in Table \ref{tab:pvalues}. All models yield statistically consistent results, with $p \approx 0.23$. We have confirmed that even large variations in the assumed template models do not significantly affect the estimated $p$-values, indicating that the results are robust to uncertainties in the theoretical modeling.

\begin{table}[h!]
\centering
\begin{tabular}{|c|c|c|}
\hline
$\nu$ & Optimal estimate (using Eq. \eqref{eq:opt_corr_estimator}) & Upper limit $\mathcal{C}^{g\,\mathrm{GW}}_{\mathrm{UL}}$ ($\ell_{\rm ref} = 8$) \\ \hline
-2 & 0.003432 & 0.00897 \\ \hline
-0.5   & 0.003203 & 0.00836 \\ \hline
0  & 0.003062 & 0.00821 \\ \hline
0.5  & 0.002995 & 0.00806 \\ \hline
\end{tabular}
\caption{Summary of the upper limits on the cross-correlation spectrum from the optimal cross-correlation estimator for different theoretical galaxy--GW correlation models. All models yield $p \approx 0.23$, indicating no statistically significant detection of the cross-correlation signal in the current dataset.}
\label{tab:pvalues}
\end{table}

 This result provides the first empirical upper bound on the correlation amplitude between the anisotropic SGWB and the large-scale galaxy distribution. Future datasets with improved sky coverage and sensitivity from both PTAs and deep galaxy surveys will significantly tighten this limit and may lead to the first detection of such a cross-correlation signal. The SKA is expected to provide nearly an order-of-magnitude improvement in timing-residual precision \citep{lazio2013square}, and with the discovery of thousands of millisecond pulsars in the SKA era \citep{smits2009pulsar}, the measurement sensitivity is anticipated to improve by roughly two orders of magnitude \citep{sah2024discovering}. Likewise, more complete galaxy catalogs will further enhance the SNR of the cross-correlation measurement.

\section{Conclusion}\label{sec:Conclusion}

In this work, we have constructed the first full analysis pipeline \texttt{PygxGW-PTA} capable of measuring the cross-correlation between pulsar-timing–based GW anisotropy and galaxy surveys. The pipeline includes optimal timing-residual cross-correlation measurements, maximum-likelihood reconstruction of the SGWB anisotropy map, computation of partial-sky coverage corrected $C_\ell$ estimator, and an optimal estimator that combines the cross-spectrum multipoles using their full noise covariance. This constitutes the first practical implementation of the theoretical formalism for galaxy--GW cross-correlation, enabling its direct application to observational data. Using this framework, we performed the first observational search for the galaxy--GW cross-correlation by jointly analyzing the \texttt{NANOGrav} 15-year dataset and the \texttt{DESI} galaxy catalog. After reconstructing the SGWB anisotropy map, we evaluated the statistical significance of the measured cross-spectrum using Monte Carlo simulations that incorporate PTA noise properties. We constructed an optimal cross-correlation estimator by taking a weighted average of the multipole moments ($\ell$), where each cross-correlation mode is weighted by the full noise covariance and by theoretical signal templates corresponding to different galaxy–GW correlation models. Using this estimator, we find no statistically significant evidence for a correlation, with a $p$-value of approximately $0.23$ across all template models considered.

Despite the absence of a detection, the analysis presented here provides the first empirical constraint on the spatial association between the nHz SGWB and the large-scale galaxy distribution. The current sensitivity is largely limited by the small number of precisely timed pulsars, anisotropic pulsar sky distribution, and the relatively large timing-residual uncertainties in existing PTA datasets. The upcoming SKA era is expected to revolutionize this landscape by providing an order-of-magnitude improvement in timing precision \citep{lazio2013square} and expanding the PTA to thousands of millisecond pulsars \citep{smits2009pulsar}. These improvements, along with the deeper, more complete maps of the galaxy with future galaxy surveys, are expected to significantly enhance the sensitivity of the cross-correlation measurement, potentially enabling a first detection of the galaxy–GW correlation.

\section*{Acknowledgments}
This work is a part of the $\langle \texttt{data|theory}\rangle$ \texttt{Universe-Lab}, which is supported by the TIFR and the Department of Atomic Energy, Government of India. This research is supported by the Prime Minister Early Career Research Award, Anusandhan National Research Foundation, Government of India. The authors express gratitude to the system administrator of the computer cluster of \texttt{⟨data|theory⟩ Universe Lab}. The authors would also like to acknowledge the use of the following Python packages in this work: Numpy \citep{van2011numpy,2020NumPy-Array}, Scipy \citep{jones2001scipy,virtanen2020scipy}, Matplotlib \citep{hunter2007matplotlib}, Astropy \citep{robitaille2013astropy,price2018astropy}, Healpy \citep{zonca2020healpy}, PINT \citep{luo2021pint,susobhanan2024pint}, Pymaster \citep{alonso2019unified}, and Pandas \citep{mckinney2011pandas}.

\bibliographystyle{mnras}
\bibliography{main}

\appendix

\section{Derivation of the optimal estimator of the angular cross-correlation}
\label{sec:Der_opt}

Our goal is to construct an optimal estimator for the angular cross-correlation,
$C_{\ell}^{\rm gGW}$, between galaxy distribution and gravitational wave background (GWB), by combining the multipoles with $\ell$-dependent weights
$Q_\ell$ in such a way that the signal-to-noise ratio (SNR) is maximized.  
We define the estimator
\begin{equation}
    \hat{S} \equiv \sum_{\ell} \hat{C}_{\ell}^{\rm gGW}\, Q_\ell
    = (\mathbf{C}^{\rm gGW})^{T} Q,
    \label{eq:opt_initial}
\end{equation}
where $\mathbf{C}^{\rm gGW}$ and $Q$ are column vectors containing
$\hat{C}_{\ell}^{\rm gGW}$ and $Q_\ell$, respectively.

The expectation value of $\hat{S}$ is
\begin{equation}
    \bar{S} \equiv \langle \hat{S} \rangle 
    = \langle (\mathbf{C}^{\rm gGW})^{T} Q \rangle
    = (\mathbf{C}^{\rm gGW}_t)^{T} Q,
    \label{eq:Mean_initial}
\end{equation}
where $\mathbf{C}^{\rm gGW}_t = \langle \mathbf{C}^{\rm gGW} \rangle$ denotes the true underlying cross-power spectrum.

The variance of $\hat{S}$ is
\begin{equation}
    \sigma^2 \equiv 
    {\rm Var}(\hat{S})
    = \sum_{\ell,\ell'} 
        Q_\ell\, \Sigma_{\rm gGW}(\ell,\ell')\, Q_{\ell'}
    = Q^{T}\, \Sigma_{\rm gGW}\, Q ,
    \label{eq:Var_initial}
\end{equation}
where $\Sigma_{\rm gGW}$ is the covariance matrix of the cross-correlation estimator $\hat{C}_{\ell}^{\rm gGW}$.

To further simplify the algebra, we follow the general strategy used in \cite{anholm2009optimal}, where an inner-product formalism is used to express the optimal-filtering problem in a compact form. We define an inner product between two vectors $X$ and $Y$ as
\begin{equation}
    [X,Y] \equiv X^{T}\, \Sigma_{\rm gGW}\, Y.
\end{equation}
This simplifies Eqs.~(\ref{eq:Mean_initial}) and (\ref{eq:Var_initial}) to
\begin{align}
    \bar{S} &= [\,\Sigma_{\rm gGW}^{-1}\mathbf{C}_t^{\rm gGW},\, Q\,],
    \label{eq:Mean_inner} \\
    \sigma^{2} &= [\,Q, Q\,].
    \label{eq:Var_inner}
\end{align}

The signal-to-noise ratio (SNR) of the estimator is
\begin{equation}
    {\rm SNR} 
    = \frac{\bar{S}}{\sigma}
    = \frac{
        [\,\Sigma_{\rm gGW}^{-1}\mathbf{C}_t^{\rm gGW},\, Q\,]}{\sqrt{[\,Q,Q\,]}}
    \label{eq:SNR_general}
\end{equation}

Using the Cauchy--Schwarz inequality for this inner product, it is straightforward to show that the SNR is maximized when
\begin{equation}
    Q = \mathcal{K} ~\Sigma_{\rm gGW}^{-1}\, \mathbf{C}_t^{\rm gGW},
\label{eq:Q}
\end{equation}
where $\mathcal{K}$ is some normalization constant. We choose $\mathcal{K}$ such that the expectation value $\bar{S}$ corresponds to the cross-correlation amplitude at a reference multipole $\ell_{\rm ref}$, i.e.,

\begin{equation}
 \mathcal{K}  = \frac{\mathbf{C}_t^{gGW}(\ell_{\rm ref})}
    {(\mathbf{C}_t^{gGW})^{\mathrm{T}}\,
    \Sigma_{\mathrm{gGW}}^{-1}\,
    \mathbf{C}_t^{\mathrm{gGW}}}.
\label{eq:K}
\end{equation}

Substituting Eqs. \eqref{eq:Q} and \eqref{eq:K} into the definition of $\hat{S}$ (Eq.~\ref{eq:opt_initial}) gives the optimal statistic
\begin{equation}
    \hat{S}_{\rm opt}
    = (\hat{\mathbf{C}}^{\rm gGW})^{T}
      \Sigma_{\rm gGW}^{-1}
      \mathbf{C}_t^{\rm gGW} ~ \Bigg[\frac{\mathbf{C}_t^{gGW}(\ell_{\rm ref})}
    {(\mathbf{C}_t^{gGW})^{\mathrm{T}}\,
    \Sigma_{\mathrm{gGW}}^{-1}\,
    \mathbf{C}_t^{\mathrm{gGW}}}\Bigg].
    \label{eq:opt_final}
\end{equation}
This represents the optimal estimator for detecting the galaxy--GWB angular cross-correlation given the covariance structure of the multipoles.

\section{Population of the SMBHB} \label{sec:pop}

The SGWB energy density per unit frequency and per unit solid angle can be written as \citep{phinney2001practical,christensen2018stochastic}
\begin{equation}
    \Omega_{\rm GW}(f, \hat{n}) = \frac{1}{\rho_c c^2} \int dz \frac{dV}{d\omega dz} \int \prod_i d\theta_i \left[ \kappa(z, \Theta_n, f_r) n_v(z, \hat{n}) \right] \left[ \frac{1}{4\pi d_L^2(z)} \frac{dE_{\rm gw}(f, \Theta_n)}{dt_r} \right],
\end{equation}
where $f_r$ is the source-frame GW frequency, $d_L(z)$ is the luminosity distance at redshift $z$, $\omega$ is the solid angle. The integration extends over the comoving volume element $dV$, and the binary-parameter set $\Theta_{\rm n} = \{\theta_i\}_{i=1}^{n}$, which includes quantities such as component mass, eccentricity, and orbital inclination.  The quantity ${dE_{\rm gw}(f_r,\Theta_{\rm n})}/{dt_r}$ corresponds to the GW power emitted by a binary with parameters $\Theta_{\rm n}$, while $n_v(z,\hat{n})$ gives the galaxy number density per unit comoving volume in the sky direction $\hat{n}$. The function $\kappa(\Theta_{\rm n}, f_r \mid z)$ encodes the average number of SMBHBs hosted by a galaxy at redshift $z$ with parameters $\Theta_{\rm n}$ emitting in a logarithmic frequency interval around $f_r$.

The quantity $\kappa$ may be expressed as
\begin{equation}
    \begin{aligned}
        \kappa(M_{\rm BH}, q, f_r|z)
        \propto
        & \int dM_{*}\, P(M_{*}|z)\, 
        P(M_{\rm BH}, q, f_r|M_{*}, z), \\
        =
        & \int dM_{*}\, P(M_{*}|z)\, P(M_{\rm BH}|M_*) P(q|M_*)\, 
        \left[P(f_r|M_*)\, f_r \right],
    \end{aligned}
    \label{pop}
\end{equation}
where $M_{\rm BH}$ is the primary black-hole mass, $q$ is the mass ratio, and $M_{*}$ refers to the stellar mass of the host galaxy. The distribution $P(M_{*}|z)$ is taken to be the galaxy stellar mass function, which we describe using a redshift-dependent Schechter form following \citet{mcleod2021evolution}. The SMBH mass distribution conditioned on host stellar mass is modeled as a log-normal distribution,
\begin{equation}
    P(M_{\rm BH}|M_{*})
    \propto 
    \frac{1}{M_{\rm BH}}
    \exp\!\Bigg[
        -\frac{ 
        (\log_{10}[M_{\rm BH}] - 
         \log_{10}[\tilde{M}_{\rm BH}(M_{*})])^{2}
        }{2\sigma_m^{2}}
    \Bigg],
\end{equation}
where the mean relation is parameterized as
\begin{equation}
    \log_{10}(\tilde{M}_{\rm BH}/M_{\odot})
    =
    \eta + \rho\, 
    \log_{10}(M_{*}/10^{11}\, M_{\odot}) + \nu ~ z,
    \label{MBH1}
\end{equation}
where $\eta$, $\rho$, and $\nu$ describe the normalization, slope of the SMBH–host relation, and redshift evolution of the relation.

The mass-ratio distribution is taken to follow
\begin{equation}
     P(q|M_{*}) \propto 
     \begin{cases}
     1/q, & 0.01 < q < 1,\\[4pt]
     0, & \text{otherwise},
     \end{cases}
     \label{q}
\end{equation}
consistent with commonly adopted parametrizations in the literature.

Throughout this analysis, we assume circular SMBHB orbits for simplicity. Under this assumption, binaries radiate at a single characteristic GW frequency in the source frame. The corresponding frequency distribution is modeled as
\begin{equation}
    P(f_r|M_{*}) \propto f_{r}^{\alpha},
    \label{freq_dist}
\end{equation}
where the exponent $\alpha$ governs the relative weight of different GW frequencies in the population.

\end{document}